\documentclass[twocolumn, aps, prl]{revtex4-2}

\usepackage{graphicx}
\usepackage{dcolumn}
\usepackage{bm}

\usepackage{amsfonts}
\usepackage{amssymb}
\usepackage{amsmath}
\usepackage{calc}
\usepackage{graphicx}
\usepackage{bm}
\usepackage{verbatim}
\usepackage{float}
\usepackage{hyperref}
\usepackage{xcolor}
\usepackage[normalem]{ulem}

\newcommand{\newtext}[1]{#1}
\newcommand{\oldtext}[1]{}

\def\be{ \begin{equation} }
\def\ee{ \end{equation} }
\def\bea{ \begin{align} }
\def\eea{ \end{align} }
\def\bse{ \begin{subequations} }
\def\ese{ \end{subequations} }

\def\ket#1{\vert #1 \rangle}

\def\Im{\,\text{Im}\,}
\def\Re{\,\text{Re}\,}

\begin{document}








\title{Twinned Dynamical Decoupling}


\author{Nayden P. Nedev}
\affiliation{Center for Quantum Technologies, Department of Physics, Sofia University, 5 James Bourchier Boulevard, 1164 Sofia, Bulgaria}

\author{Nikolay V. Vitanov}
\affiliation{Center for Quantum Technologies, Department of Physics, Sofia University, 5 James Bourchier Boulevard, 1164 Sofia, Bulgaria}



\date{\today}

\begin{abstract}
Systematic pulse-area errors limit the fidelity of quantum control across many qubit platforms. 
We introduce twinned dynamical decoupling (TDD), an analytic family of sequences \(T2n\) in which a pulse sequence is paired with its $\pi$-phase-shifted twin. 
This $\pi$-phase step cancels common-mode systematic pulse-area errors to all orders on exact resonance.
Then the phases of the pulses in each of the constituent twins are determined in such a manner that detuning errors are suppressed to the highest possible order as well.
We have derived a simple analytic formula for these phases applicable to arbitrary sequence length. 
We demonstrate the sequences with superconducting transmon qubits on the IBM Quantum processor ibm\_torino and the IQM Quantum processor Garnet. 
The measured population plateaus agree closely with theory and show enhanced robustness compared to the most frequently used dynamical decoupling protocols. 
\end{abstract}


\maketitle

Dynamical decoupling (DD) is one of the standard tools for protecting quantum
systems from decoherence. 
Its basic idea is to apply a sequence of control pulses that refocuses unwanted evolution and averages out the effect of the environment. 
In this way, coherence can be extended without measurements, feedback, or the overhead of full quantum error correction~\cite{Viola,viola1999,Viola2005,Khodjasteh2005,Khodjasteh2007,DLidar,Yang2011,Suter2016}. 
Originally developed in magnetic resonance, DD is now widely used in quantum information processing, quantum sensing, and precision spectroscopy, and has been demonstrated on a variety of platforms, including trapped ions, ultracold atoms, superconducting qubits, solid-state spins, and photonic systems~\cite{Alvarez2010,Peng2011, Biercuk2009, Damodarakurup2009, Lucamarini2011, Wang2016, DDSupercond, Pokharel2024,Barthel2023,Xu2025XiSwap,Tong2025}.
Its practical importance stems from the fact that DD can often be implemented with the same elementary controls already available for quantum gates, memories, and sensing protocols.

Many DD protocols have been developed for different noise models and hardware
constraints. The Carr--Purcell--Meiboom--Gill (CPMG) sequence  \cite{Carr,Meiboom}, Uhrig
dynamical decoupling (UDD) \cite{Uhrig}, the Knill dynamical decoupling (KDD) family \cite{Souza}, and the XY-type sequences \cite{Maudsley,CDD, QDD} all improve coherence by exploiting suitable pulse timings, pulse
phases, or symmetry properties. These sequences have proved extremely useful
in suppressing dephasing noise and stabilizing qubit evolution. However, their
performance in real devices is not determined only by the noise that acts
between the pulses. It also depends crucially on the accuracy of the pulses
themselves.
%
A common imperfection is a pulse-area error, caused,
for example, by amplitude miscalibration, imperfect pulse duration, or slow
drift of the drive strength. Another common imperfection is detuning, which may arise from frequency miscalibration, ac-Stark shifts, or slow drift of the
qubit transition frequency. Such errors are often systematic during a single
experimental shot. They therefore do not behave as random noise, but can
accumulate coherently over the sequence. As a result, adding more pulses may
improve the filtering of environmental noise while at the same time increasing
the sensitivity to pulse imperfections.

This tradeoff is especially relevant for present-day quantum processors,
where coherent calibration errors and slow drifts are often among the dominant
limitations. A useful DD sequence should therefore do more than suppress
unwanted free evolution: it should also be robust against systematic errors in
the pulses used to implement it. Standard sequences achieve this only
partially. CPMG works very well in selected settings but is sensitive to
pulse-area errors outside its natural operating conditions. XY and KDD-type
sequences improve the compensation of certain pulse errors through phase
symmetrization. UDD optimizes the pulse timings for dephasing noise. Yet these
constructions do not, in general, provide exact cancellation of systematic
pulse-area errors to all orders, nor do they give a simple analytic phase
prescription that can be extended to arbitrary sequence length while also
improving detuning robustness.

In this work, we introduce twinned dynamical decoupling (TDD), an analytic family of DD sequences denoted by T2$n$. 
The construction is based on a simple idea. 
We take a palindromic sequence of nominal $\pi$ pulses and then apply its twin, in which all pulse phases are shifted by $\pi$. 
This phase pairing makes the full sequence exactly insensitive to systematic pulse-area errors on resonance. 
The cancellation is not merely perturbative: on exact resonance, the propagator is the identity for an
arbitrary systematic deviation of the pulse area, as long as the same deviation applies to all pulses in the sequence.

Once this all-order pulse-area cancellation is built into the structure of the sequence, the remaining pulse phases are chosen to suppress detuning errors. 
We show that this can be done analytically. 
In particular, we derive a closed-form expression for the phases that is valid for arbitrary sequence length. Thus, the sequences require no numerical optimization and can be constructed directly for any desired number of pulses. 
Increasing the sequence length then increases the order to which detuning errors are suppressed.

We demonstrate the T$2n$ sequences with superconducting transmon qubits on the IBM Quantum processor ibm\_torino and the IQM Quantum processor Garnet and compare them with the most used DD protocols.

\oldtext{The derivation of the DD sequences is presented in the Supplemental Material \cite{supplemental}.}
\newtext{The essential derivation is presented here; further algebraic details and the arbitrary-order proof are given in the Supplemental Material \cite{supplemental}.}
The target DD propagator is the unit matrix $\mathbf{I}$.
\newtext{A single phased pulse is represented by
\be
U_{\phi}=\left[\begin{matrix}a&be^{i\phi}\\-b^*e^{-i\phi}&a^*\end{matrix}\right],
\ee
with $|a|^2+|b|^2=1$, and the propagator of an $n$-pulse block is
$U^{(n)}=U_{\phi_n}\cdots U_{\phi_2}U_{\phi_1}$.}
The sequences feature the general twin structure 
\be\label{T2n}
T2n = (S^{(n)})_0 (S^{(n)})_\pi.
\ee
\newtext{Here $(S^{(n)})_\pi$ is obtained from $(S^{(n)})_0$ by shifting every pulse phase according to $\phi_k\mapsto\phi_k+\pi$, without changing the pulse shapes, pulse areas, durations, or delays.}
The twin component $S_0$ may contain an odd or even number of nominal (i.e., in the absence of errors) phased $\pi$ pulses
\footnote{A phased pulse $\theta_\phi$ corresponds on resonance to the rotation 
$R_\phi(\theta)=\exp\left[-\frac{i\theta}{2} (\sigma_x\cos\phi-\sigma_y\sin\phi)\right]$.}
with delays $\tau$ between the pulses, and has the palindrome structure
\be
S^{(n)}:\ \tau/2 - \pi_{\phi_1} - \tau - \pi_{\phi_2} - \tau - \cdots - \tau - \pi_{\phi_2} - \tau - \pi_{\phi_1} - \tau/2.
\ee
\oldtext{This palindrome symmetry is necessary in order to ensure the complete cancellation of the pulse area errors on exact resonance by the total sequence $T2n$.}
\newtext{This palindrome symmetry is a sufficient condition for the diagonal Cayley--Klein parameter of the block propagator to be real on resonance. Pairing this block with its $\pi$-shifted twin then produces the identity for an arbitrary common pulse-area error.}

Having obtained in this manner complete cancellation of the pulse area errors on resonance,  
\oldtext{the phases in each twin sequence $S^{(n)}$ are derived by maximizing the performance of $T2n$ against the detuning. It is shown in the Supplemental Material \cite{supplemental} that this task reduces to maximizing the robustness of the imaginary part of the matrix element $U^{(n)}_{11}$ of the propagator induced by $S^{(n)}$.}
\newtext{the phases in each twin block $S^{(n)}$ are chosen to suppress the error of the full $T_{2n}$ propagator against detuning. This criterion can be stated exactly.}
Indeed, the matrix elements of the total propagator read
\be
U^{(2n)}_{11} = 1 + 2i U^{(n)}_{11} \Im U^{(n)}_{11}, \quad 
U^{(2n)}_{12} = 2i U^{(n)}_{12} \Im U^{(n)}_{11} ,
\ee
and hence the Frobenius distance $\epsilon$ between the actual propagator and the target unit matrix $\mathbf{I}$ is
\be\label{fidelity}
\epsilon = 2\sqrt{2}\, |\Im U^{(n)}_{11}|.
\ee
The standard average gate fidelity for small error is given by $\mathcal{F}=1-\epsilon^2/3 +O(\epsilon^4)$.
Our task is therefore defined as: find palindrome composite sequences $S^{(n)}$ which minimize $|\Im U^{(n)}_{11}|$ against the detuning $\Delta$ to as many orders in $\Delta$ as possible.
We note that no conditions apply to the real part $\Re U^{(n)}_{11}$ or the off-diagonal part $U^{(n)}_{12}$.
\newtext{In other words, the phases are selected so that the largest possible number of successive Taylor coefficients of $\Im U^{(n)}_{11}$ vanish at zero detuning. }

\begin{table}[tb]
\renewcommand{\arraystretch}{1.3}
  \centering
  \begin{tabular}{|c|l|c|}
     \hline
     T${2n}$ & Phases of T${2n}$ & $\Im U^{(n)}_{11}$ \\ \hline
     T${2}$ & $(0, 1)\pi$ & $\Delta$ \\
     T${4}$ & $(0, 0, 1, 1)\pi$ & $\frac\pi 2 \Delta^3$ \\
     T${6}$ & $(0, \frac{1}{3},0,1, \frac{4}{3}, 1)\pi$ & $\Delta^3$ \\
     T${8}$ & $(0, \frac{1}{2}, \frac{1}{2},0,1, \frac{3}{2}, \frac{3}{2}, 1)\pi$ & $\pi\Delta^5$ \\
     T${10}$ & $(0, \frac{3}{5}, \frac{4}{5}, \frac{3}{5}, 0, 1, \frac{8}{5}, \frac{9}{5}, \frac{8}{5}, 1)\pi$ & $\Delta^5$ \\
     T${12}$ & $(0, \frac{2}{3}, 1, 1, \frac{2}{3}, 0, 1, \frac{5}{3}, 0, 0, \frac{5}{3}, 1)\pi$ & $\frac32\pi\Delta^7$ \\
     T${14}$ & $(0, \frac{5}{7}, \frac{8}{7}, \frac{9}{7}, \frac{8}{7}, \frac{5}{7}, 0, 1, \frac{12}{7}, \frac{1}{7}, \frac{2}{7}, \frac{1}{7}, \frac{12}{7}, 1)\pi$ & $\Delta^7$ \\
     T${16}$ & $(0, \frac{3}{4}, \frac{5}{4}, \frac{3}{2}, \frac{3}{2}, \frac{5}{4}, \frac{3}{4}, 0, 1, \frac{7}{4}, \frac{1}{4}, \frac{1}{2}, \frac{1}{2}, \frac{1}{4}, \frac{7}{4}, 1)\pi$ & $2\pi\Delta^9$ \\
     T${18}$ & $(0, \frac{7}{9}, \frac{4}{3}, \frac{5}{3}, \frac{16}{9}, \frac{5}{3}, \frac{4}{3}, \frac{7}{9}, 0, $ & \\
     & \qquad $ 1, \frac{16}{9}, \frac{1}{3}, \frac{2}{3}, \frac{7}{9}, \frac{2}{3}, \frac{1}{3}, \frac{16}{9}, 1)\pi$ & $\Delta^9$ \\
     T${20}$ & $(0, \frac{4}{5}, \frac{7}{5}, \frac{9}{5}, 0, 0, \frac{9}{5}, \frac{7}{5}, \frac{4}{5}, 0, $ & \\
     & \qquad $ 1, \frac{9}{5}, \frac{2}{5}, \frac{4}{5}, 1, 1, \frac{4}{5}, \frac{2}{5}, \frac{9}{5}, 1)\pi$ & $\frac52\pi\Delta^{11}$ \\
     T${22}$ & $(0, \frac{9}{11}, \frac{16}{11}, \frac{21}{11}, \frac{2}{11}, \frac{3}{11}, \frac{2}{11},\frac{21}{11}, \frac{16}{11}, \frac{9}{11}, 0,$ & \\
     & \qquad $1, \frac{20}{11}, \frac{5}{11}, \frac{10}{11},\frac{13}{11}, \frac{14}{11}, \frac{13}{11}, \frac{10}{11}, \frac{5}{11}, \frac{20}{11},  1)\pi$ & $\Delta^{11}$ \\
     T${24}$ & $(0, \frac{5}{6}, \frac{3}{2}, 0, \frac{1}{3}, \frac{1}{2}, \frac{1}{2}, \frac{1}{3}, 0, \frac{3}{2}, \frac{5}{6}, 0,$ & \\
     & \qquad $1, \frac{11}{6}, \frac{1}{2}, 1, \frac{4}{3}, \frac{3}{2}, \frac{3}{2}, \frac{4}{3}, 1, \frac{1}{2}, \frac{11}{6}, 1)\pi$ & $3\pi\Delta^{13}$ \\
     \hline
   \end{tabular}
  \caption{Phases of several high-order T$2n$ sequences.
  The last column presents the leading term in the detuning $\Delta$ of $\Im U^{(n)}_{11}$, which determines the Frobenius error.}
  \label{table:Tn}
\end{table}

There are multiple solutions for each number of pulses $n$, with similar performance.
Here we present one of these, the phases of which are given by the formula
\be\label{phases}
\phi_k = \frac{(k-1)(n-k)}{n}\pi,\quad k=1,2,\ldots,2n,
\ee
which is valid for any number of pulses $n$.

The explicit values are given in Table \ref{table:Tn}.
The last column lists the first nonzero term in the Taylor expansion of $\Im U^{(n)}_{11}$ versus the detuning, which, according to Eq.~\eqref{fidelity}, determines the first nonzero term of the Frobenius error.
These follow the pattern (see Supplemental Material at \cite{supplemental})
\bse
\begin{align}
\epsilon_{2n} &= 2\sqrt{2}\, |\Delta|^n +O(|\Delta|^{n+1}),\quad n\ \text{odd}, \\
\epsilon_{2n} &= 2\sqrt{2}\,\frac{n\pi} 4 |\Delta|^{n+1}+O(|\Delta|^{n+2}),\quad n\ \text{even}.
\end{align}
\ese

In this respect, given the exact analytic formula for the composite phases \eqref{phases}, we can claim that detuning errors can be suppressed to \textit{any} order.

\newtext{
The TDD phase pattern and the switching function play complementary roles. Consider a pure-dephasing interaction $H_{\mathrm{SB}}(t)=\frac12\beta_z(t)\sigma_z$.
Every ideal equatorial $\pi$ pulse reverses $\sigma_z$, independently of its azimuthal phase,
$R_\phi(\pi)^\dagger\sigma_zR_\phi(\pi)=-\sigma_z$.
In the toggling frame of the control, the interaction therefore becomes
$\widetilde H_{\mathrm{SB}}(t)=\frac12 y(t)\beta_z(t)\sigma_z$, with $y(t)=(-1)^{m(t)}$,
where $m(t)$ is the number of completed $\pi$ pulses. For symmetric timing, $\int_0^{T_{\mathrm{seq}}}y(t)\,dt=0$, and hence a static or quasistatic dephasing field is refocused exactly in the ideal-pulse limit. For time-dependent dephasing, $y(t)$ defines the usual echo-type filter function. Thus, the repeated inversions suppress environmental dephasing, whereas the TDD phases make these inversions robust against systematic pulse imperfections. This argument establishes dephasing suppression; it does not by itself imply universal decoupling of an arbitrary three-axis system-bath interaction.
}





For the demonstrations on real quantum hardware, we have used two quantum processor units (QPUs) of IBM Quantum and IQM. 
In the first set, we have tested our DD sequences on qubit 10 of ibm{\_}torino, one of IBM Quantum Heron r1 Processors. 
It is an open-access quantum processor that consists of 133 transmon qubits. 
The demonstrations were performed using $R_X(\theta)$ fractional rotation gates \cite{Frac_gate}, each of duration $32$ ns, which allowed efficient implementation of $R_X(\theta)$ for an arbitrary rotation angle $\theta$, avoiding the decomposition into multiple native gates. 
This allowed us to test an arbitrarily large deviation from an ideal $\pi$-pulse rotation (X gate) without relying on low-level control ({see also Supplemental Material} \cite{supplemental}). 

The demonstration consists of preparing the excited state $\ket{1}$ by applying an $X$ gate to the ground state of the qubit $\ket{0}$. Then a DD sequence is applied by $R_X(\theta)$ with the corresponding phases added by $R_Z(\phi)$ gates and symmetrical time delays between the pulses. 
The DD sequence is followed by another $X$ gate and finally a measurement in the computational basis. 
Each circuit is repeated 512 times, and the results are averaged to obtain a single data point for our plots.

\begin{figure}[tb]
\includegraphics[width=0.9\columnwidth]{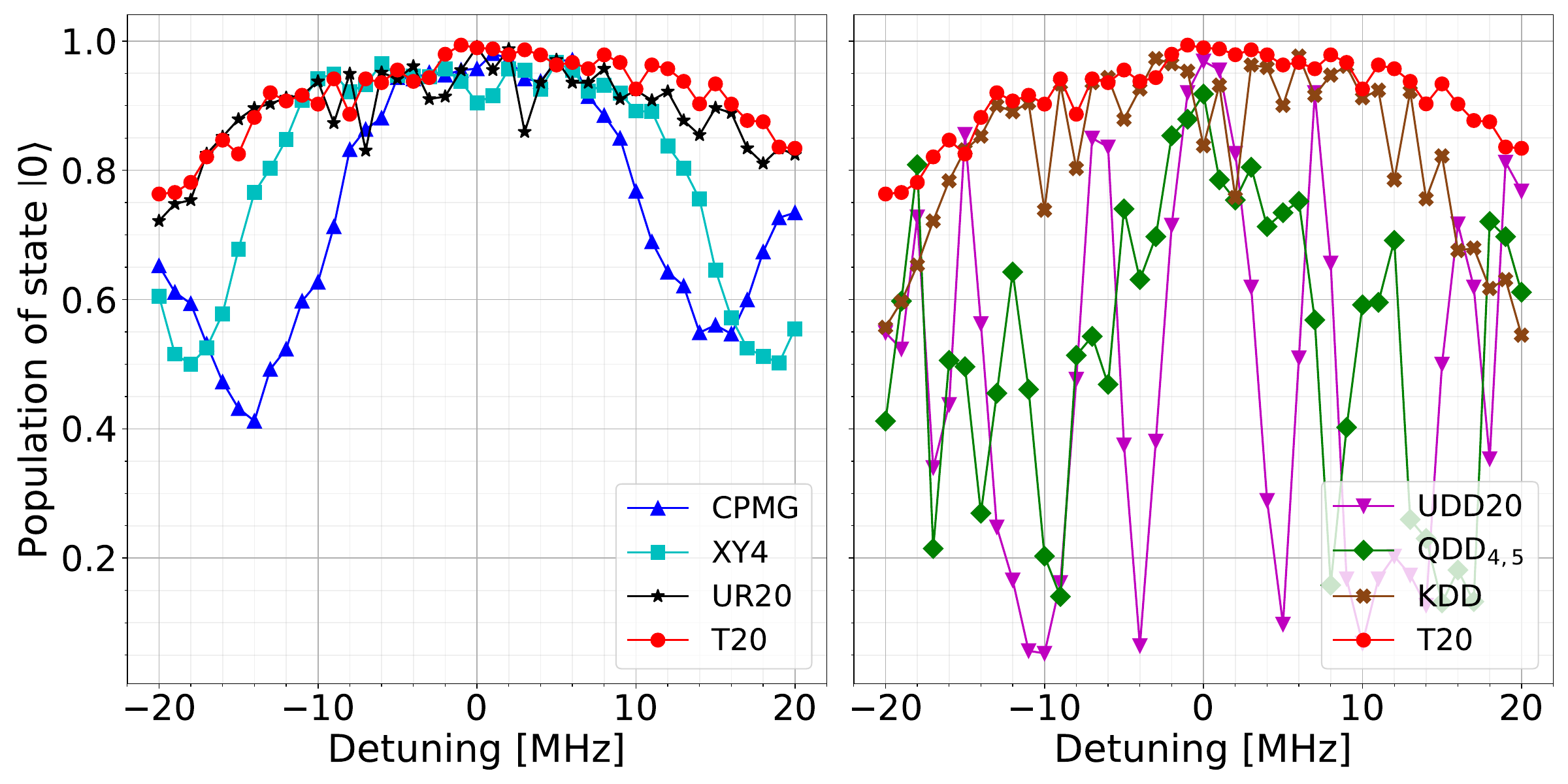} 
\caption{Measured results on IQM Garnet, comparing some of the most widely used types of DD sequences against detuning.}
\label{exp_results_comp_detun}
\end{figure}

The other demonstrations are done using qubit 1 of IQM Garnet, a 20-qubit quantum processing unit based on superconducting transmon qubits. 
The demonstrations presented here are performed using the pulse-level access quantum computing module IQM Pulla \cite{iqm_pulla}. 
Here, we implemented the DD sequences using the native PRX gates with their calibrated "drag{\_}crf" implementation. 
Each pulse duration was $20$ ns, and the pulse amplitude and drive frequency were varied. 
Again, each circuit was repeated 512 times, and the results were averaged to obtain a single data point for our plots.

\begin{figure*}[tbph]
\includegraphics[width=1.8\columnwidth]{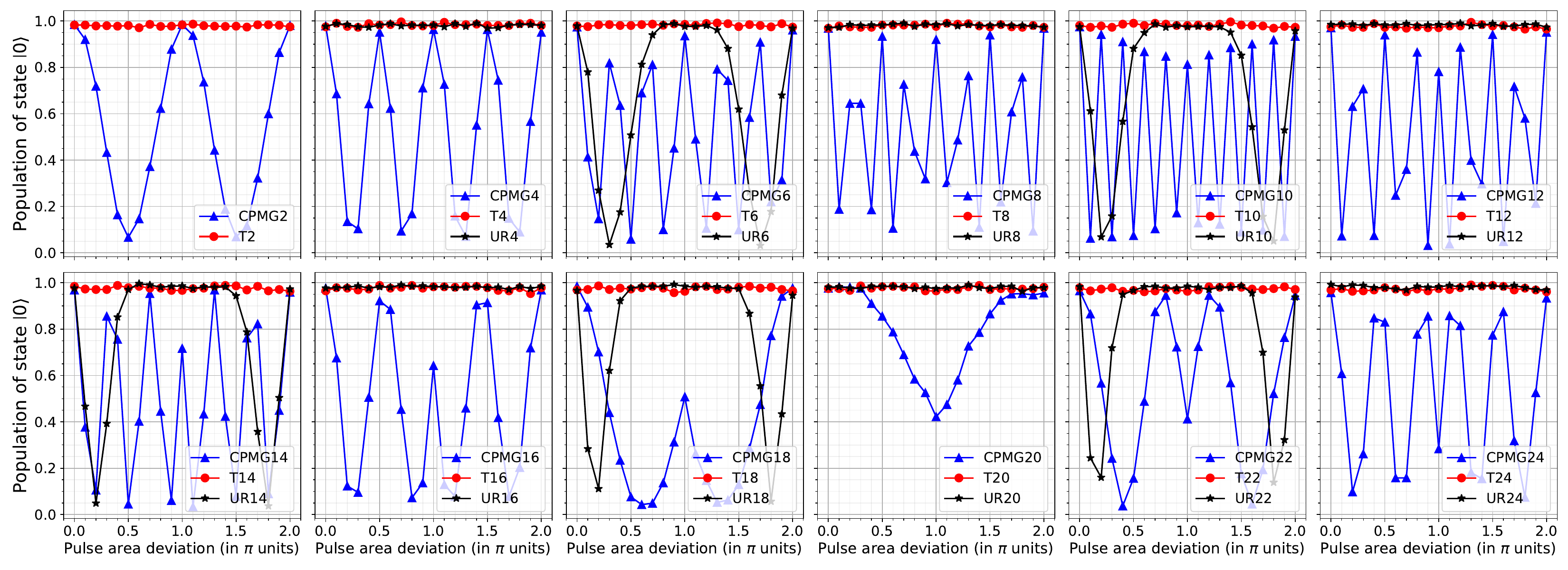}
\caption{Measured $\ket{0}$-state population for several T$2n$ sequences on ibm{\_}torino compared with the UR$2n$ sequences and the respective CPMG$2n$ sequences with the same number of pulses.}
\label{exp_data}
\end{figure*}

Figure \ref{exp_data} presents the measured results for the $T2n$ sequences of 2 to 24 pulses as a function of the pulse area, along with the results for the CPMG and UR$2n$ \cite{URDD} sequences with the same number of pulses.
The T$2n$ sequences of any number of pulses compensate for arbitrarily large pulse area errors, as predicted.
The UR$2n$ sequences perform similarly for even $n$ (with 4, 8, 12, ... pulses), but hold only in a finite pulse area range for odd $n$, which expands with $n$ but remains finite.
The repeated CPMG sequences are very sensitive to pulse area errors, for any number of repetitions $n$.

\begin{figure*}[tbph]
\includegraphics[width=1.8\columnwidth]{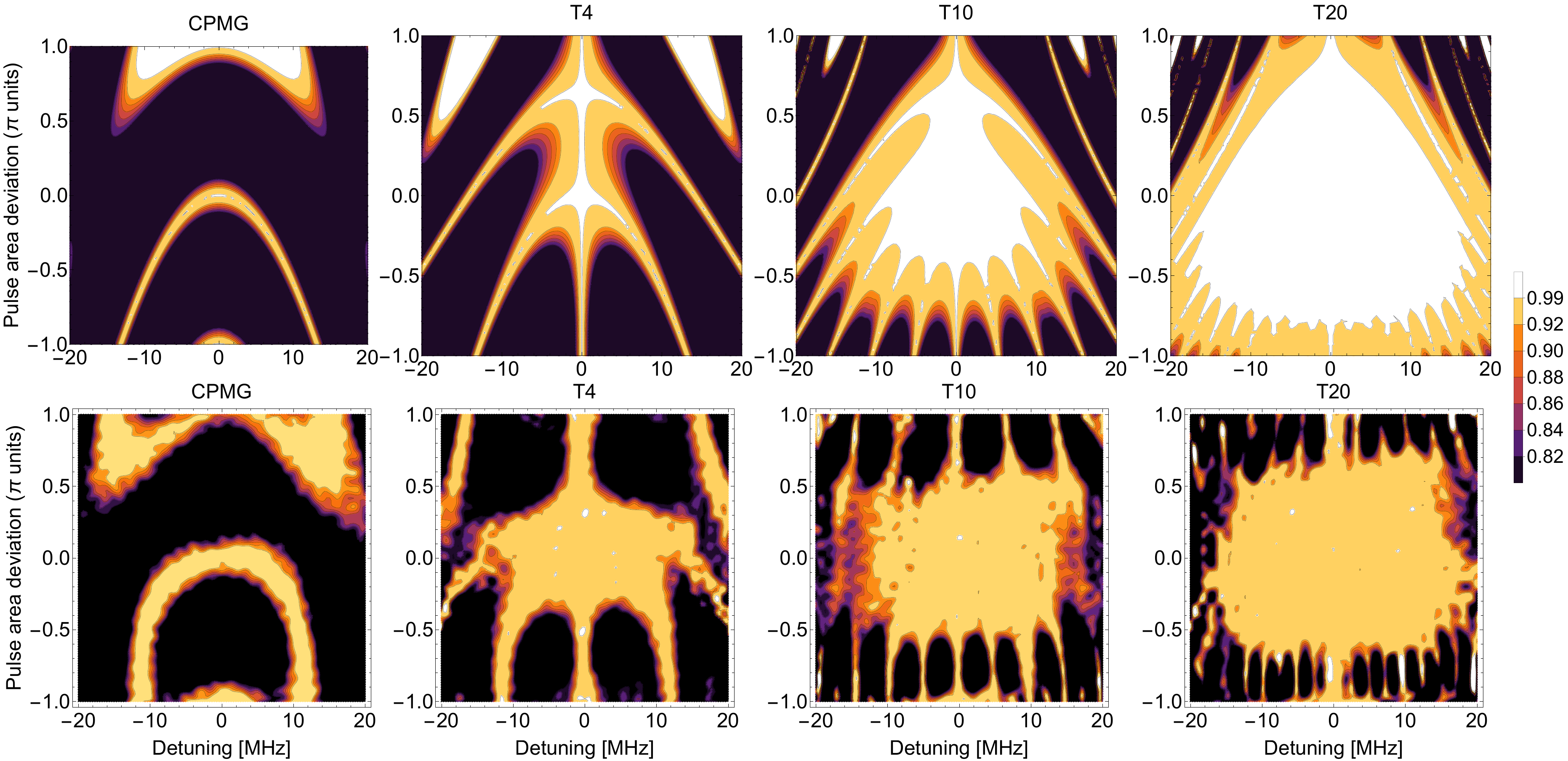}
\caption{Theoretical predictions (top row) and measured results from IQM Garnet (bottom row) of the $\ket{0}$-state population for several T\(2n\) sequences and CPMG with simultaneous deviations in both detuning and pulse area. The Rabi frequency is $\Omega =2\pi\times 25$ MHz and the pulse duration is 20 ns.}
\label{2D_exp_results}
\end{figure*}

\begin{figure}[tb]
\includegraphics[width=0.9\columnwidth]{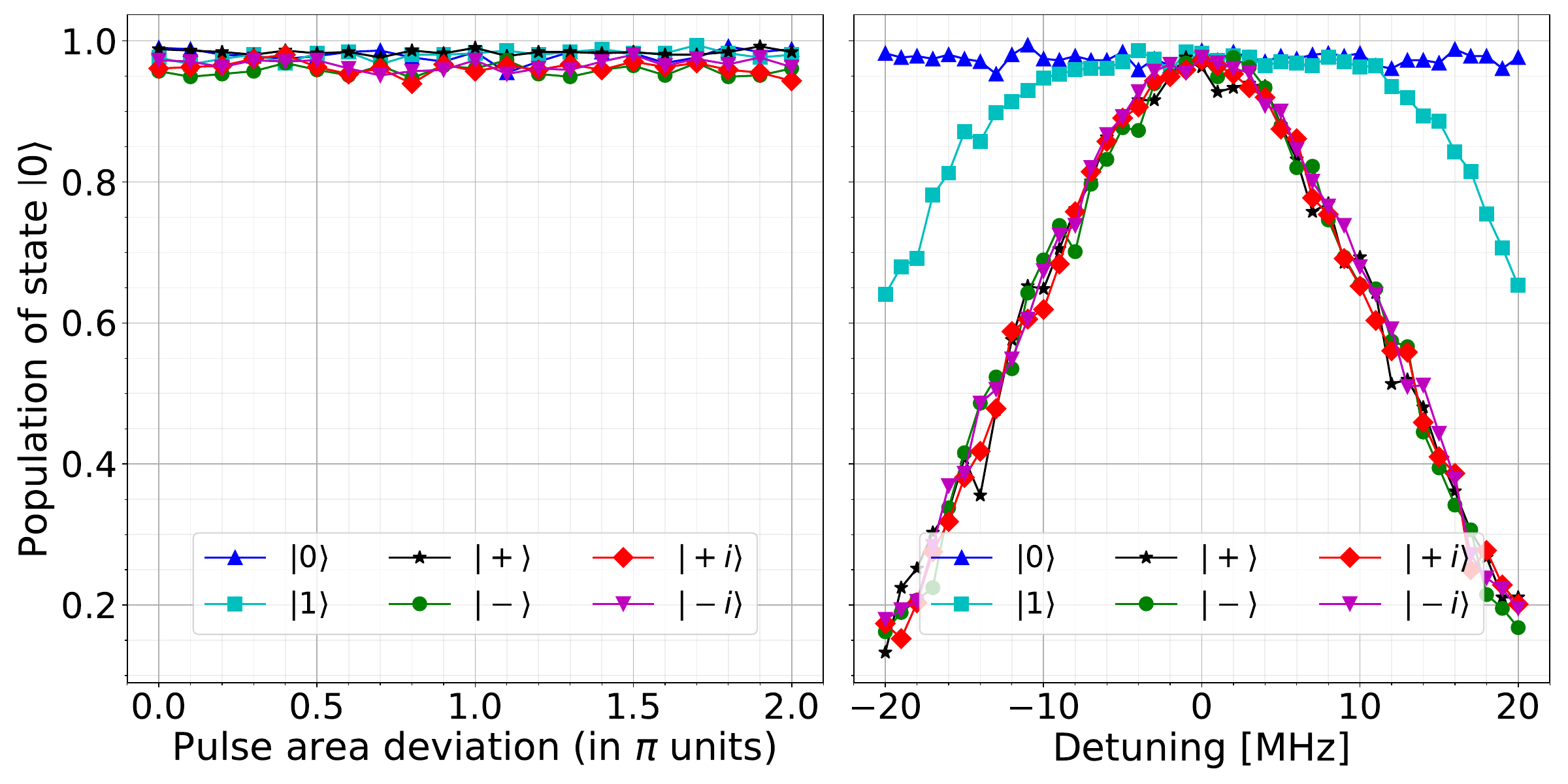} 
\caption{Measured results on ibm{\_}torino and IQM Garnet of T10 robustness for different initial states.}
\label{exp_results_diff_initial_states}
\end{figure}

\begin{figure}[tbph]
\begin{tabular}{cc}
\includegraphics[width=0.9\columnwidth]{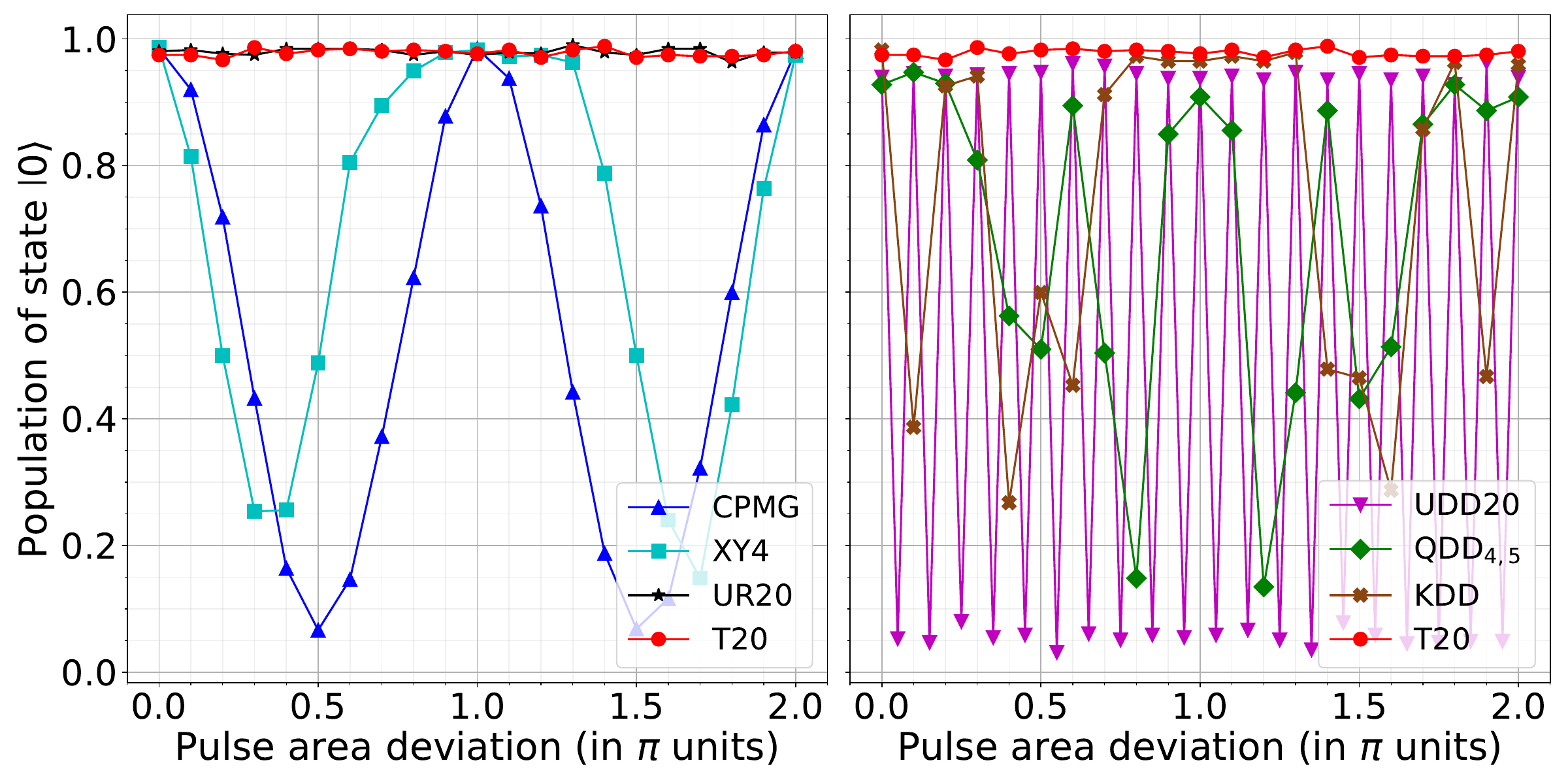} \\
\includegraphics[width=0.9\columnwidth]{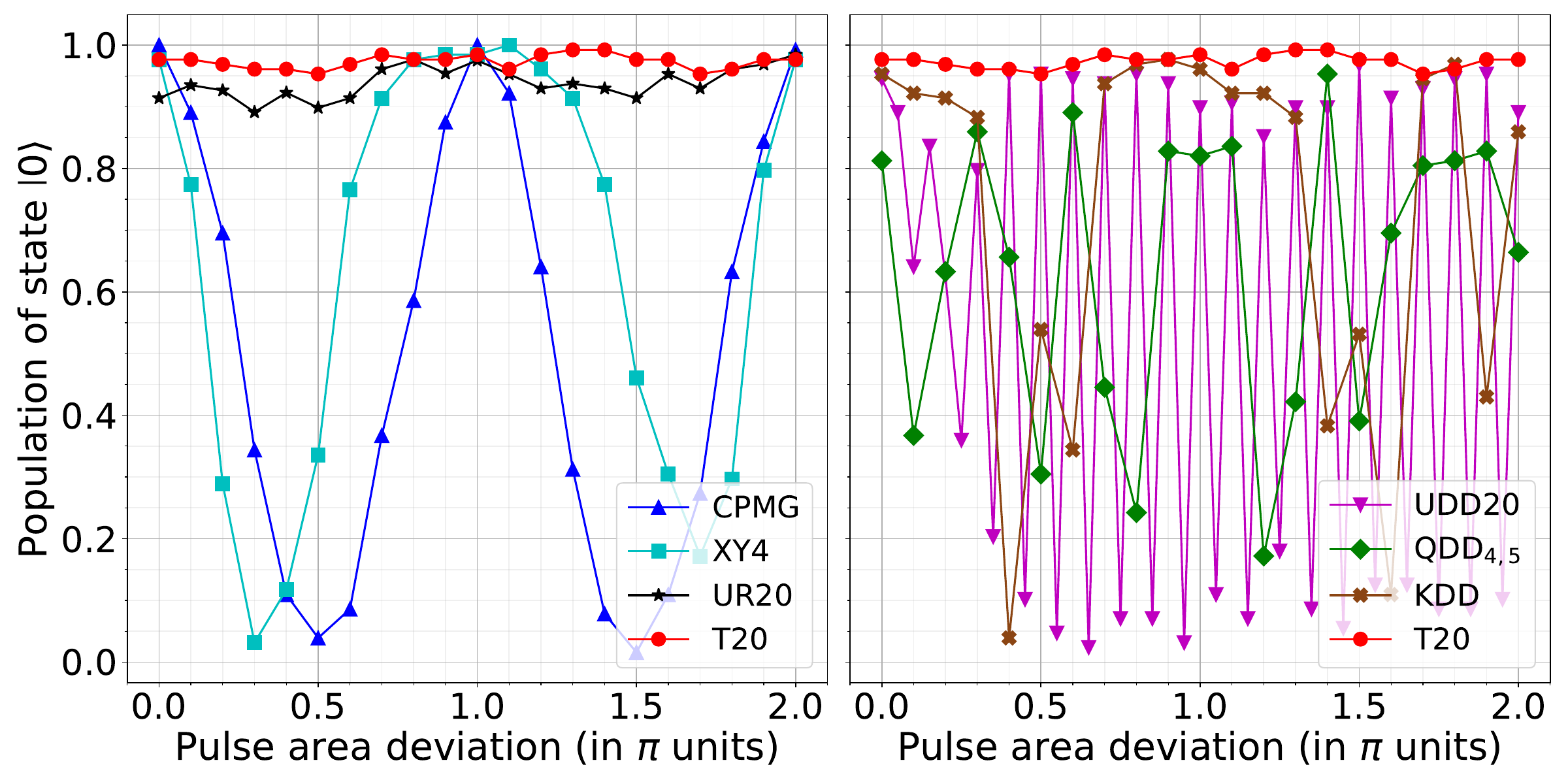}
\end{tabular}
\caption{Measured results on ibm{\_}torino (top row) and IQM Garnet (bottom row), comparing some of the most widely used types of DD sequences against systematic pulse-area errors. The T20 results (red dots) are plotted in both frames of each row for the sake of comparison.}
\label{exp_results_comp_torino}
\end{figure}

In the next demonstration, Fig. \ref{exp_results_comp_torino} compares the performance of the T$2n$ sequences with the most widely used types of DD sequences. 
Moreover, we show measurement results from both IBM (top row) and IQM (bottom row) QPUs, with remarkably similar performance.
The T20 sequence clearly outperforms all others by a large margin, with only UR20 matching it on IBM's QPU (but UR20 is clearly inferior on IQM).
Among the other DD sequences, we note the XY4 and KDD sequences, which perform reasonably well.

Figure \ref{exp_results_comp_detun} compares these DD sequences against detuning. 
The measurements were performed on IQM's QPU because it provides pulse level control.
Although the performance of T20 is not as perfect as against the pulse area error in the preceding figures, it comfortably outperforms all other DD sequences, including UR20, which comes closest to it.
Here KDD comes third, then XY4, CPMG, QDD and UDD.

Figure \ref{2D_exp_results} shows the measured results on the IQM QPU for the simultaneous deviations in both the pulse area and the detuning 
for CPMG and several T$2n$ sequences. 
The data in the bottom row in general confirm the theoretical predictions on the top row.
All T$2n$ sequences outperform the CPMG sequence, as expected.
Moreover, the high-fidelity domain expands from T4, through T10, to T20, as predicted.
The deviations from the theoretical predictions are most likely dominated by state preparation and measurement (SPAM) errors, which limit the observed ground-state probability in our measurements.

Finally, we have tested the performance of the T10 sequence for different initial states of the qubit against pulse area (on IBM) and detuning (on IQM) errors.
This demonstration follows the setup previously described; however, initially, the six Pauli eigenstates are prepared as $\ket{\Psi}=R\ket{0}$, where $R$ is the appropriate unitary rotation. 
Then a DD sequence is added, after that we apply the reverse operation $R^{\dagger}$, and finally measure in the computational basis.

The results are shown in Figure \ref{exp_results_diff_initial_states}.
\oldtext{As the left panel demonstrates, the complete pulse error compensation of T10 is independent of the initial state. However, the results differ with respect to detuning errors. Complete error cancellation occurs for the initial state $\ket{0}$, and a broad cancellation range is observed for state $\ket{1}$ too. However, the detuning robustness decreases for the superposition initial states $\ket{+}$, $\ket{-}$, $\ket{+i}$, and $\ket{-i}$. This behavior is consistent with the presence of dephasing processes, which affect the superposition states only.}
\newtext{The broad pulse-area robustness of T10 is observed for all six Pauli eigenstates, as expected for an identity operation. Under detuning, the measured population is more robust for the computational-basis states than for the equatorial states. This state dependence is consistent with a residual rotation about the $z$ axis, which leaves the populations of $\ket{0}$ and $\ket{1}$ unchanged but changes the phases of their superpositions; stochastic dephasing can cause an additional reduction of equatorial-state coherence. Population measurements alone do not distinguish these two effects.}

The demonstrations on the cloud-access QPUs of IBM and IQM show that the T$2n$ sequences outperform by a large margin the most popular previous DD sequences, with only UR coming close (but still below) for sequences with a number of pulses that are multiples of 4.
\oldtext{Our sequences are designed to suppress systematic errors in the pulse area and the detuning. As in all previous DD sequences, arbitrary pulse-to-pulse errors cannot be canceled exactly. In other words, our sequences are designed to compensate \textit{slow} noise. This is the usual situation in experiments, as our demonstration also shows. Our sequences demand that no drifts in the experimental parameters occur during the sequence itself, so that the ingredient pulses are identical (except for their phases), albeit imperfect. Our sequences can compensate for slow drift that may make each shot different from the others, but during the shot, no drift should occur. This assumption, which is indeed made in all previously known DD sequences, is quite natural because in the QPU demonstrations, the DD sequence in each shot lasted less than 1 $\mu$s, and we do not expect noise changes on such a scale. Importantly, Ramsey coherence measurements on the superconducting quantum processors demonstrate that the proposed sequences retain the fundamental ability to suppress environmental decoherence while providing enhanced robustness against systematic control errors ({see also Supplemental Material} \cite{supplemental}).}
\newtext{The exact all-order pulse-area cancellation requires the pulse-area error to be common to all constituent pulses during one sequence. It therefore applies to calibration offsets and drifts that are slow compared with the sequence duration, but it does not imply exact cancellation of arbitrary pulse-to-pulse fluctuations. This common-mode control-error assumption must be distinguished from the environmental noise being filtered: environmental dephasing need not be strictly static, and its suppression is governed by the switching function and the associated filter function. Ramsey measurements on both processors demonstrate an extension of coherence relative to free evolution and show that this protection persists when controlled pulse-area or detuning errors are introduced (see Supplemental Material \cite{supplemental}). On ibm{\_}torino, T10 increases the fitted coherence time from $T_2^*=238.3~\mu$s to $T_2=344.6~\mu$s; on IQM Garnet, it increases from $T_2^*=10.1~\mu$s to $T_2=43.9~\mu$s.}

In conclusion, we have introduced a new analytic family of twinned dynamical decoupling sequences — \(T2n\) — that achieve exact cancellation of systematic pulse-area errors to all orders.
The construction is rooted in simple conditions imposed on the pulse phases, leading to closed-form analytic expressions for an arbitrary-order sequence. 
We demonstrate these sequences on superconducting transmon qubits from both IBM Quantum processor ibm{\_}torino and IQM Quantum processor Garnet, observing population plateaus in close agreement with theory. 
We also demonstrate superior performance to the most widely used types of DD sequences, with only UR showing comparable results, as expected from the theoretical predictions. 
These results open a new path for hardware-efficient quantum control, especially in architectures where systematic calibration drift is a limiting factor. 
We anticipate applications in quantum memories, quantum sensing, and error-protected gate synthesis, as well as integration into hybrid error mitigation and correction frameworks.

This research is supported by the Bulgarian Ministry of Education and Science under the National Program “Young Scientists and Postdoctoral Students – 2”, by the Bulgarian national plan for recovery and resilience, Contract No. BG-RRP-2.004-0008-C01 (SUMMIT), Project No. 3.1.4, and by the European Union’s Horizon Europe research and innovation program under Grant Agreement No. 101046968 (BRISQ). 
We acknowledge the use of the supercomputing cluster PhysOn at Sofia University for this work. 
We acknowledge the use of IBM Quantum services for this work. The views expressed are those of the authors and do not reflect the official policy or position of IBM or the IBM Quantum team. 
We acknowledge the use of IQM Resonance services for this work. The views expressed are those of the authors and do not reflect the official policy or position of IQM.

\bibliographystyle{apsrev4-2}
\bibliography{references} 

\end{document}






\title{Supplemental Material to Twinned Dynamical Decoupling}




\author{Nayden P. Nedev}
\affiliation{Center for Quantum Technologies, Department of Physics, Sofia University, 5 James Bourchier Boulevard, 1164 Sofia, Bulgaria}

\author{Nikolay V. Vitanov}
\affiliation{Center for Quantum Technologies, Department of Physics, Sofia University, 5 James Bourchier Boulevard, 1164 Sofia, Bulgaria}


\date{\today}

\maketitle

\section{Derivation of the TDD sequences}
The dynamics of a qubit due to an external pulsed interaction is described by a propagator $\U$, which is conveniently parameterized with the complex Cayley-Klein parameters $a$ and $b$ ($|a|^2+|b|^2=1$) as
\begin{equation}\label{U}
\U=\begin{bmatrix}
    a & b\\
    -b^\ast & a^\ast
\end{bmatrix}.
\end{equation}
For exact resonance ($\Delta = 0$), the Schrödinger equation is solved exactly for any $\Omega(t)$. Then the Cayley-Klein parameters depend only on the pulse area $A = \int_{t_i}^{t_f}\Omega(t)dt$:
$a = \cos(A/2)$, $b = -i \sin(A/2)$,
with $\Omega(t)$ assumed real. 
The transition probability is $p = |b|^2 = \sin^2(A/2)$, and hence, complete population inversion occurs for $A = \pi$ ($\pi$ pulses) or odd-integer multiples of $\pi$. This inversion is sensitive to variations in the pulse area: a small deviation from the value $\pi$, i.e., $A = \pi(1 + \xi)$, causes an error in the inversion of order $\mathcal{O}(\xi^2)$,
\begin{equation}
p = 1 - \pi^2 \xi^2/4 + \mathcal{O}(\xi^4).
\end{equation}

A constant phase shift $\phi$ in the Rabi frequency, $\Omega(t)\to \Omega(t)e^{i\phi}$, is imprinted into the propagator (\ref{U}) as
\begin{equation}\label{Uphi}
\U(\phi) =
\begin{bmatrix}
a & b e^{i\phi} \\
-b^\ast e^{-i\phi} & a^\ast
\end{bmatrix}.
\end{equation}
A sequence of $N$ pulses, each with a phase $\phi_k$, produces the propagator
\begin{equation}
\U^{(N)} = \U(\phi_N) \U(\phi_{N-1}) \cdots \U(\phi_1).
\end{equation}

The objective of any DD sequence is to produce the identity propagator,
\be
\I = \left[\begin{array}{cc}
1 & 0 \\
0 & 1
\end{array}\right] ,
\ee
as closely as possible, i.e., by compensating deviations in the relevant parameters to the highest possible order.

We define the DD error $\epsilon$ as the Frobenius distance between the actual propagator $\U$ and the target propagator $\I$:
\be
\epsilon = || \mathbf{U} - \mathbf{I}|| = \sqrt{2|U_{11} - 1|^2 + 2|U_{12}|^2}.
\ee

We begin with the observation that a sequence of two pulses with the same interaction parameters, but with a phase shift of $\pi$ in the second Rabi frequency, is described by the overall propagator
\be\label{topology}
\U(\pi) \U(0) = 
\begin{bmatrix}
a^2+|b|^2 & b (a-a^\ast) \\
b^\ast (a-a^\ast) & (a^\ast)^2+|b|^2
\end{bmatrix}.
\ee
If $a$ is real, then we obtain 
\be
\U(\pi) \U(0) = \mathbf{I},\quad \text{if}\ a \in \mathbb{R}.
\ee
This simple observation constitutes the main argument of the present paper: No matter what the interaction is, if its structure is described by the twinned propagator of Eq.~\eqref{topology}, then a perfect restoration of the initial state occurs, i.e., a perfect dynamical decoupling.

The next question is: when is the Cayley-Klein parameter $a$ real? 
The first obvious answer is: on exact resonance. Then $a=\cos(A/2)$, as it is well known.
The second, less obvious case is off resonance under specific symmetry conditions: the Rabi frequency must be a symmetric function of time, $\Omega(-t)=\Omega(t)$, and the detuning must be anti-symmetric, $\Delta(-t)=-\Delta(t)$ \cite{Vitanov1999}.

Now, we want to replace the two propagators in Eq.~\eqref{topology} by sequences of pulses.
We have to answer the question: When does the overall element $U_{11}$ of a composite sequence remain real?
The answer is: for symmetric sequences, e.g., by sequences of the same pulses with palindrome phases $\{ \phi_1, \phi_2, \ldots, \phi_2,\phi_1\}$.
The propagator for a palindrome sequence, for even and odd numbers of pulses in it, can be written as
\bse\label{overall}
\begin{align}
&\U(\phi_1)\cdots[\U(\phi_{n-2})[\U(\phi_{n-1})\U(\phi_n)\U(\phi_n)\U(\phi_{n-1})] \notag \\
&\quad  \times \U(\phi_{n-2})]\cdots \U(\phi_1),\\
&\U(\phi_1)\cdots[\U(\phi_{n-2})[\U(\phi_{n-1})\U(\phi_n)\U(\phi_{n-1})] \U(\phi_{n-2})] \notag\\ 
&\quad\cdots \U(\phi_1).
\end{align}
\ese
Because we have
\be\label{phiphi}
\U(\phi) \U(\phi) = 
\begin{bmatrix}
a^2-|b|^2 & b e^{i\phi} (a+a^\ast) \\
-b^\ast e^{-i\phi} (a+a^\ast) & (a^\ast)^2-|b|^2
\end{bmatrix},
\ee
we conclude that for both even and odd numbers of pulses in the calculation of the overall propagator, we have the repeated concatenated blocks of the kind
\be\label{phichiphi}
\U(\phi) \mathbf{V}(\chi) \U(\phi),
\ee
where $\mathbf{V}(\chi)$ has the same structure as $\U(\phi)$, but with different parameters
\begin{equation}\label{Uphi1}
\mathbf{V}(\chi) =
\begin{bmatrix}
c & d e^{i\chi} \\
-d^\ast e^{-i\chi} & c^\ast
\end{bmatrix}.
\end{equation}
The multiplication in Eq.~\eqref{phichiphi} for real $a$ and $c$ gives for the top left element the expression
\be
(a^2-|b|^2) c -2a \Re (b d^\ast e^{i(\phi-\chi)}),
\ee
which is real-valued.
The execution of the multiplication in the overall propagator in Eq.~\eqref{overall} keeps the top left element of the propagator real after each step, hence we obtain real diagonal elements in the end as well.

Consequently, the desired unit propagator $\mathbf{I}$ is obtained \textit{exactly} if (i) the DD sequence is composed of two identical parts (twins), one of which is shifted by phase $\pi$ compared to the other, Eq.~\eqref{topology}, (ii) if each of these parts is a sequence of pulses with palindrome phases, Eqs.~\eqref{overall}, and (iii) if the Cayley-Klein parameter $a$ for every single pulse is real. 
The latter condition (iii) is fulfilled on exact resonance and for the special case of chirped-pulse excitation where the Rabi frequency is symmetric, and the detuning is anti-symmetric in time.
Therefore, sequences that satisfy conditions (i) and (ii) above, which we will assume hereafter, provide a perfect DD on resonance, i.e., they compensate errors in the pulse area $A$ to an infinite order.

Now, having assured perfect DD on resonance, we turn our attention to the performance of the DD sequences off resonance. 
Our objective is to find DD sequences that provide as large a compensation as possible with respect to the detuning.
We start from Eq.~\eqref{topology}, which gives the overall propagator as a function of the parameters of the two parts that form the composite sequence.
Because the target is the unit propagator, we conclude that we should have 
\bse\label{cond}
\begin{align}
a^2+|b|^2 &= 1 +2 i a \Im a \approx 1, \\
b(a-a^\ast) &= 2ib\Im(a) \approx 0. 
\end{align}
\ese
The Frobenius distance between the actual propagator and the target unit matrix is
\be
\epsilon = \sqrt{2 [4|a|^2 (\Im a)^2 + 4|b|^2 (\Im a)^2]} = 2\sqrt{2}\, |\Im a|.
\ee

Therefore, our task is redefined as: find palindrome composite sequences which minimize $|\Im a|$ and maximize robustness against the detuning.
We note that no conditions apply to the real part, $\Re(a)$.


The derivation of the twin DD sequences consists in expanding $\Im a$ in a Taylor series versus the detuning $\Delta$ and canceling as many terms in the ascending order as possible.
In other words, we generate a set of equations from the conditions $\Im[\partial^{k}_{\Delta}\U^{(n)}_{11} ]_{\Delta=0}=0$ with $k=0,1,2,\ldots$. 
Note that only the \textit{imaginary} parts need to be annulled.

The \(T2n\) sequences can be divided into two subclasses of even and odd numbers of pulses $n$ in each of the twin blocks. (i) The sequences $T{4l}$ of $4l$
pulses $(n=2l)$, where $l=1,2,\ldots$, have the structure
\begin{equation}
T_{4l} = (r_l)_0 (r_l^{-1})_0 (r_l)_\pi (r_l^{-1})_\pi ,
\end{equation}
where
\begin{align}
(r_l)_0 &=
(\phi_1,\phi_2,\ldots,\phi_{l-1},\phi_l), \\
(r_l^{-1})_0 &=
(\phi_l,\phi_{l-1},\ldots,\phi_2,\phi_1), \\
(r_l)_\pi &=
(\phi_1+\pi,\phi_2+\pi,\ldots,\phi_{l-1}+\pi,\phi_l+\pi), \\
(r_l^{-1})_\pi &=
(\phi_l+\pi,\phi_{l-1}+\pi,\ldots,\phi_2+\pi,\phi_1+\pi).
\end{align}

The sequences \(T2n\) of $2n$ pulses, where $n=2j+1$ and
$j=0,1,2,\ldots$, have the structure
\begin{equation}
T_{2n} = (r_n)_0 (r_n)_\pi ,
\end{equation}
where
\begin{align}
(r_n)_0 &=
(\phi_1,\phi_2,\ldots,\phi_{n-1},\phi_n), \\
(r_n)_\pi &=
(\phi_1+\pi,\phi_2+\pi,\ldots,\phi_{n-1}+\pi,\phi_n+\pi).
\end{align}

There are multiple solutions for the phases, with similar performance. We list
here one set of solutions, with the general formula for their phases,
\begin{equation}
\phi_k =
\frac{(k-1)(n-k)}{n}\pi,
\quad
(k=1,2,\ldots,2n).
\end{equation}




\section{Derivation of the leading-order detuning scaling}

We prove the leading-order behavior quoted in Eq.~(6) in the main text for the phases
\begin{equation}
\phi_k=\frac{(k-1)(n-k)}{n}\pi,
\quad k=1,2,\ldots,2n .
\label{eq:phase_formula_proof}
\end{equation}
It is sufficient to consider one twin block,
\begin{equation}
S^{(n)}=U_{\phi_n}U_{\phi_{n-1}}\cdots U_{\phi_1},
\end{equation}
because the full twinned sequence gives the Frobenius error
\begin{equation}
\epsilon_{2n}
=
2\sqrt{2}\,
\left|\operatorname{Im}U^{(n)}_{11}\right|,
\label{eq:DF_ImU11}
\end{equation}
where \(U^{(n)}\) is the propagator of one block \(S^{(n)}\).

For a nominal \(\pi\)-pulse with small dimensionless detuning \(\Delta\), we write
\begin{equation}
U_\phi(\Delta)
=
\begin{pmatrix}
a & b e^{i\phi}\\
-b^*e^{-i\phi} & a^*
\end{pmatrix},
\end{equation}
with
\begin{equation}
a=C-iY ,
\end{equation}
where
\begin{equation}
C=-\frac{\pi}{4}\Delta^2+O(\Delta^4),
\quad
Y=\Delta+O(\Delta^3),
\label{eq:CY_expansion}
\end{equation}
and
\begin{equation}
|b|=1+O(\Delta^2).
\end{equation}
Thus, the imaginary part of \(a\) starts at order \(\Delta\), whereas the real part starts only at order \(\Delta^2\).

The phase formula \eqref{eq:phase_formula_proof} implies
\begin{equation}
\phi_{k+1}-\phi_k
=
\frac{n-2k}{n}\pi .
\label{eq:phase_difference}
\end{equation}
Therefore, the phase factors appearing in the toggling-frame expansion form a complete set of \(n\)-th roots of unity. Consequently, all lower powers in the detuning expansion cancel. Equivalently, this follows from the finite-root identity
\begin{equation}
\prod_{m=0}^{n-1}
\left(1+z e^{2\pi i m/n}\right)
=
1-(-z)^n .
\label{eq:root_identity}
\end{equation}
Equivalently, the lower-order coefficients are elementary
symmetric polynomials of the complete set of \(n\)-th roots of
unity and therefore vanish for all orders below \(n\).

First set \(C=0\), i.e. \(a=-iY\). Then, the diagonal element of the one-block propagator has the leading structure
\begin{equation}
U^{(n)}_{11}
=
\begin{cases}
(-1)^{(n+1)/2}iY^n+O(Y^{n+2}),
& n \ \text{odd},\\[4pt]
(-1)^{n/2}+O(Y^2),
& n \ \text{even}.
\end{cases}
\label{eq:block_Czero}
\end{equation}
For odd \(n\), the first surviving imaginary contribution is therefore proportional to \(Y^n\). For even \(n\), the block remains real at \(C=0\), and hence
\begin{equation}
\operatorname{Im}U^{(n)}_{11}=0,
\quad
C=0,\quad n \ \text{even}.
\end{equation}

We now restore the small real part \(C\). For even \(n\), the leading imaginary term is obtained by replacing one of the \(n\) factors \(-iY\) by \(C\). Since there are \(n\) equivalent choices, one obtains
\begin{equation}
\operatorname{Im}U^{(n)}_{11}
=
(-1)^{n/2} n C Y^{n-1}
+
O(CY^{n+1},C^2).
\label{eq:even_C_correction}
\end{equation}
Substituting Eq.~\eqref{eq:CY_expansion}, we find, for odd \(n\),
\begin{equation}
\operatorname{Im}U^{(n)}_{11}
=
(-1)^{(n+1)/2}\Delta^n
+
O(\Delta^{n+1}),
\end{equation}
and therefore
\begin{equation}
\left|\operatorname{Im}U^{(n)}_{11}\right|
=
|\Delta|^n
+
O(|\Delta|^{n+1}),
\quad n \ \text{odd}.
\label{eq:odd_ImU}
\end{equation}
For even \(n\), Eq.~\eqref{eq:even_C_correction} gives
\begin{equation}
\operatorname{Im}U^{(n)}_{11}
=
(-1)^{n/2+1}
\frac{n\pi}{4}
\Delta^{n+1}
+
O(\Delta^{n+2}),
\end{equation}
and hence
\begin{equation}
\left|\operatorname{Im}U^{(n)}_{11}\right|
=
\frac{n\pi}{4}
|\Delta|^{n+1}
+
O(|\Delta|^{n+2}),
\quad n \ \text{even}.
\label{eq:even_ImU}
\end{equation}

Finally, using Eq.~\eqref{eq:DF_ImU11}, we obtain
\begin{equation}
\epsilon_{2n}
=
2\sqrt{2}\,
|\Delta|^n
+
O(|\Delta|^{n+1}),
\quad n \ \text{odd},
\label{eq:DF_odd}
\end{equation}
and
\begin{equation}
\epsilon_{2n}
=
2\sqrt{2}\,
\frac{n\pi}{4}
|\Delta|^{n+1}
+
O(|\Delta|^{n+2}),
\quad n \ \text{even}.
\label{eq:DF_even}
\end{equation}
Thus, Eq.~(6) in the main text follows, where \(\epsilon_{2n}\) denotes the Frobenius error, not the standard average gate infidelity.

%
%


\section{Effect of free evolution between the pulses}


In the main derivation, we first ignored the free evolution between consecutive pulses. Here, we show that symmetric free-evolution intervals do not change the order of the detuning suppression, although they may change the numerical prefactor of the leading term.

We consider a sequence in which each pulse is surrounded symmetrically by half-delays,
\begin{equation}
\tau/2-\pi_{\phi_1}-\tau-\pi_{\phi_2}-\cdots-\tau-\pi_{\phi_n}-\tau/2 .
\end{equation}
The free evolution due to a constant detuning is described by
\begin{equation}
F=
\begin{pmatrix}
e^{i\Delta\tau/2} & 0\\
0 & e^{-i\Delta\tau/2}
\end{pmatrix}.
\end{equation}
Thus, each elementary pulse propagator is replaced by
\begin{equation}
U_\phi
\longrightarrow
\widetilde U_\phi=F U_\phi F .
\label{eq:free_evolution_replacement}
\end{equation}
This replacement should be applied at the elementary-pulse level, not as a final multiplication of the total composite propagator.

For a single pulse, we write
\begin{equation}
U_\phi=
\begin{pmatrix}
a & b e^{i\phi}\\
-b^*e^{-i\phi} & a^*
\end{pmatrix}.
\end{equation}
Then Eq.~\eqref{eq:free_evolution_replacement} gives
\begin{equation}
\widetilde U_\phi
=
\begin{pmatrix}
a e^{i\Delta\tau} & b e^{i\phi}\\
-b^*e^{-i\phi} & a^* e^{-i\Delta\tau}
\end{pmatrix}.
\end{equation}
Therefore, the free evolution does not change the transverse phase \(\phi\), but modifies the diagonal Cayley--Klein parameter according to
\begin{equation}
a\longrightarrow \widetilde a=a e^{i\Delta\tau}.
\label{eq:a_free_evolution}
\end{equation}

We now write
\begin{equation}
a=C-iY,
\end{equation}
where, for a nominal \(\pi\)-pulse and small detuning,
\begin{equation}
C=c_2\Delta^2+O(\Delta^4),
\quad
Y=y_1\Delta+O(\Delta^3).
\label{eq:CY_general}
\end{equation}
For the normalization used in the main text, one has
\begin{equation}
c_2=-\frac{\pi}{4},
\quad
y_1=1 .
\end{equation}
Using Eq.~\eqref{eq:a_free_evolution}, we obtain
\begin{equation}
\widetilde a
=
(C-iY)e^{i\Delta\tau}
=
\widetilde C-i\widetilde Y,
\end{equation}
with
\begin{align}
\widetilde C
&=
C\cos(\Delta\tau)+Y\sin(\Delta\tau),
\\
\widetilde Y
&=
Y\cos(\Delta\tau)-C\sin(\Delta\tau).
\end{align}
Substituting Eq.~\eqref{eq:CY_general}, we find
\begin{align}
\widetilde C
&=
(c_2+y_1\tau)\Delta^2+O(\Delta^4),
\label{eq:Ctilde}
\\
\widetilde Y
&=
y_1\Delta+O(\Delta^3).
\label{eq:Ytilde}
\end{align}
Hence, the free evolution preserves the parity structure of the single-pulse Cayley--Klein parameter:
\begin{equation}
\operatorname{Re}\widetilde a=O(\Delta^2),
\quad
\operatorname{Im}\widetilde a=O(\Delta).
\end{equation}

Because the phase-root cancellation of the twinned sequence depends on this parity structure and on the phases \(\phi_k\), which are not changed by the free evolution, the order of the detuning suppression remains unchanged. The only modification is in the leading prefactor for the even-\(n\) subclass.

Indeed, using the same expansion as in the zero-delay case, the one-block propagator satisfies, for odd \(n\),
\begin{equation}
\operatorname{Im}U^{(n)}_{11}
=
(-1)^{(n+1)/2}
\widetilde Y^n
+
O(\Delta^{n+1}).
\end{equation}
Therefore,
\begin{equation}
\left|\operatorname{Im}U^{(n)}_{11}\right|
=
|y_1|^n |\Delta|^n
+
O(|\Delta|^{n+1}),
\quad n \ \text{odd}.
\label{eq:free_odd}
\end{equation}
For even \(n\), the leading imaginary contribution is obtained by replacing one of the \(n\) factors proportional to \(\widetilde Y\) by the real correction \(\widetilde C\). This gives
\begin{equation}
\operatorname{Im}U^{(n)}_{11}
=
(-1)^{n/2}
n \widetilde C \widetilde Y^{n-1}
+
O(\Delta^{n+2}),
\end{equation}
and therefore
\begin{equation}
\left|\operatorname{Im}U^{(n)}_{11}\right|
=
n |c_2+y_1\tau|\, |y_1|^{n-1}
|\Delta|^{n+1}
+
O(|\Delta|^{n+2}),
\quad n \ \text{even}.
\label{eq:free_even}
\end{equation}

Finally, since the full twinned sequence has a Frobenius distance
\begin{equation}
\epsilon_{2n}
=
2\sqrt{2}
\left|
\operatorname{Im}U^{(n)}_{11}
\right|,
\end{equation}
we obtain
\begin{equation}
\epsilon_{2n}
=
2\sqrt{2}\,
|y_1|^n|\Delta|^n
+
O(|\Delta|^{n+1}),
\quad n \ \text{odd},
\end{equation}
and
\begin{equation}
\epsilon_{2n}
=
2\sqrt{2}\,
n |c_2+y_1\tau|\, |y_1|^{n-1}
|\Delta|^{n+1}
+
O(|\Delta|^{n+2}),
\quad n \ \text{even}.
\end{equation}
Thus, symmetric free evolution between the pulses does not reduce the order of detuning suppression. It can, however, change the numerical coefficient of the leading even-\(n\) term. In the zero-delay limit \(\tau=0\), using \(c_2=-\pi/4\) and \(y_1=1\), we recover
\begin{equation}
\epsilon_{2n}
=
2\sqrt{2}\,
|\Delta|^n
+
O(|\Delta|^{n+1}),
\quad n \ \text{odd},
\end{equation}
and
\begin{equation}
\epsilon_{2n}
=
2\sqrt{2}\,
\frac{n\pi}{4}
|\Delta|^{n+1}
+
O(|\Delta|^{n+2}),
\quad n \ \text{even}.
\end{equation}

\section{Proof of reality of the diagonal propagator element in a special off-resonance case}

In the main text, we use the fact that the twinned cancellation
mechanism applies whenever the diagonal Cayley--Klein parameter
\(a\) of the relevant pulse or pulse block is real. On exact
resonance, this is immediate. Here, we show that the same reality
condition also holds for a special class of off-resonant pulses
with time-reversal symmetry.

We consider a driven two-level system described in the rotating
frame by
\begin{equation}
H(t,\phi)
=
\frac{\hbar}{2}
\begin{pmatrix}
\Delta(t) & \Omega(t)e^{i\phi}\\
\Omega(t)e^{-i\phi} & -\Delta(t)
\end{pmatrix},
\label{eq:H_phi_symmetry}
\end{equation}
where \(\Omega(t)\) is real and \(\phi\) is a constant control
phase. We assume that the pulse is centered at \(t=0\) and that
the Rabi frequency and detuning obey
\begin{equation}
\Omega(-t)=\Omega(t),
\quad
\Delta(-t)=-\Delta(t).
\label{eq:even_odd_conditions}
\end{equation}
These conditions are satisfied by standard analytic models such
as the Landau--Zener and Allen--Eberly models.

The constant phase \(\phi\) can be removed by a time-independent
basis transformation. Let
\begin{equation}
R_\phi=
\begin{pmatrix}
e^{i\phi/2} & 0\\
0 & e^{-i\phi/2}
\end{pmatrix}.
\end{equation}
Then
\begin{equation}
H(t,\phi)=R_\phi H(t,0)R_\phi^\dagger,
\end{equation}
where
\begin{equation}
H(t,0)
=
\frac{\hbar}{2}
\begin{pmatrix}
\Delta(t) & \Omega(t)\\
\Omega(t) & -\Delta(t)
\end{pmatrix}.
\label{eq:H_zero_phase}
\end{equation}
Consequently, if \(U_0(T,-T)\) is the propagator generated by
\(H(t,0)\), then the propagator generated by \(H(t,\phi)\) is
\begin{equation}
U_\phi(T,-T)=R_\phi U_0(T,-T)R_\phi^\dagger .
\label{eq:U_phi_gauge}
\end{equation}
This transformation changes only the phases of the off-diagonal
matrix elements. The diagonal Cayley--Klein parameter \(a\) is
unchanged. Therefore, it is sufficient to prove the reality of
\(a\) for \(\phi=0\).

For \(\phi=0\), the Hamiltonian is real and symmetric,
\begin{equation}
H(t,0)^T=H(t,0),
\end{equation}
and the symmetry conditions \eqref{eq:even_odd_conditions} imply
\begin{equation}
H(-t,0)=\sigma_x H(t,0)\sigma_x ,
\label{eq:H_time_reversal_zero_phase}
\end{equation}
where \(\sigma_x\) is the Pauli \(x\) matrix.

We now discretize the evolution interval \([-T,T]\) symmetrically.
Let \(0<t_1<t_2<\cdots<t_N=T\), and define the elementary
propagators on the positive half of the pulse by
\begin{equation}
A_j=\exp\left[-\frac{i}{\hbar}H(t_j,0)\delta t\right],
\quad j=1,\ldots,N .
\end{equation}
Since \(H(t_j,0)\) is symmetric, each \(A_j\) is also symmetric,
\begin{equation}
A_j^T=A_j .
\end{equation}
The corresponding elementary propagators on the negative half
are, by Eq.~\eqref{eq:H_time_reversal_zero_phase},
\begin{equation}
B_j
=
\exp\left[-\frac{i}{\hbar}H(-t_j,0)\delta t\right]
=
\sigma_x A_j\sigma_x .
\end{equation}
The full propagator can therefore be written as
\begin{equation}
U_0(T,-T)
=
A_N A_{N-1}\cdots A_1
B_1 B_2\cdots B_N .
\end{equation}
Using \(B_j=\sigma_x A_j\sigma_x\) and \(\sigma_x^2=I\), this becomes
\begin{equation}
U_0(T,-T)
=
P\sigma_x P^T\sigma_x ,
\label{eq:U0_P_form}
\end{equation}
where
\begin{equation}
P=A_NA_{N-1}\cdots A_1,
\quad
P^T=A_1A_2\cdots A_N .
\end{equation}
Equation~\eqref{eq:U0_P_form} immediately gives
\begin{equation}
\sigma_x U_0(T,-T)^T\sigma_x=U_0(T,-T).
\label{eq:U0_symmetry_relation}
\end{equation}

The propagator of a two-level system may be written in
Cayley--Klein form as
\begin{equation}
U_0(T,-T)
=
\begin{pmatrix}
a & b\\
-b^* & a^*
\end{pmatrix}.
\label{eq:CK_zero_phase}
\end{equation}
Substituting Eq.~\eqref{eq:CK_zero_phase} into
Eq.~\eqref{eq:U0_symmetry_relation}, we obtain
\begin{equation}
\begin{pmatrix}
a & b\\
-b^* & a^*
\end{pmatrix}
=
\begin{pmatrix}
a^* & b\\
-b^* & a
\end{pmatrix}.
\end{equation}
Hence,
\begin{equation}
a=a^*,
\end{equation}
so the diagonal Cayley--Klein parameter is real.

Returning to a general constant phase \(\phi\), Eq.~\eqref{eq:U_phi_gauge}
gives
\begin{equation}
U_\phi(T,-T)
=
\begin{pmatrix}
a & b e^{i\phi}\\
-b^* e^{-i\phi} & a^*
\end{pmatrix},
\label{eq:CK_general_phase_real_a}
\end{equation}
with the same real \(a\). Thus, the reality of the diagonal
Cayley--Klein parameter is independent of the constant pulse
phase.

We now apply this result to the twinned construction. Consider
two propagators with phases differing by \(\pi\),
\begin{equation}
U_\phi
=
\begin{pmatrix}
a & b e^{i\phi}\\
-b^* e^{-i\phi} & a^*
\end{pmatrix},
\quad
U_{\phi+\pi}
=
\begin{pmatrix}
a & -b e^{i\phi}\\
b^* e^{-i\phi} & a^*
\end{pmatrix}.
\end{equation}
Their product is
\begin{equation}
U_{\phi+\pi}U_\phi
=
\begin{pmatrix}
a^2+|b|^2 & b e^{i\phi}(a-a^*)\\
b^*e^{-i\phi}(a-a^*) & (a^*)^2+|b|^2
\end{pmatrix}.
\end{equation}
If \(a\) is real, then the off-diagonal elements vanish and
using \(|a|^2+|b|^2=1\), we obtain
\begin{equation}
U_{\phi+\pi}U_\phi=I .
\end{equation}
Therefore, the twinned cancellation mechanism applies not only
on exact resonance, but also to the special off-resonant class
of pulses satisfying Eq.~\eqref{eq:even_odd_conditions}. This is
a symmetry-based extension and should not be interpreted as cancellation of arbitrary detuning noise.

\section{Detailed description of the experimental demonstrations}

In this section, we describe the implementation of the \(T2n\) sequences on two superconducting platforms and provide details of the Ramsey-based coherence measurements used to evaluate their robustness.
\paragraph{Timing structure of the dynamical decoupling sequences.}
Each \(T2n\) sequence consists of $2n$ nominal $\pi$ pulses separated by free-evolution intervals of equal duration $\tau$, arranged symmetrically around the center of the sequence as 
\begin{equation}
    \tau /2-\pi -\tau - \dots -\pi -\tau /2.
\end{equation}
The total sequence duration is therefore
\begin{equation}
T_{\mathrm{seq}} =2 n (t_p + \tau),
\end{equation}
where $t_p$ is the duration of the physical pulse.
For all data presented, the inter-pulse delay $\tau$ is kept fixed within each experiment, and comparisons between different sequences are performed either at fixed $\tau$ or fixed $T_{\mathrm{seq}}$, depending on the figure.

\begin{figure*}[h!t!]
\centering
\begin{minipage}[h!t!]{0.48\textwidth}
    \centering
    \includegraphics[width=\linewidth]{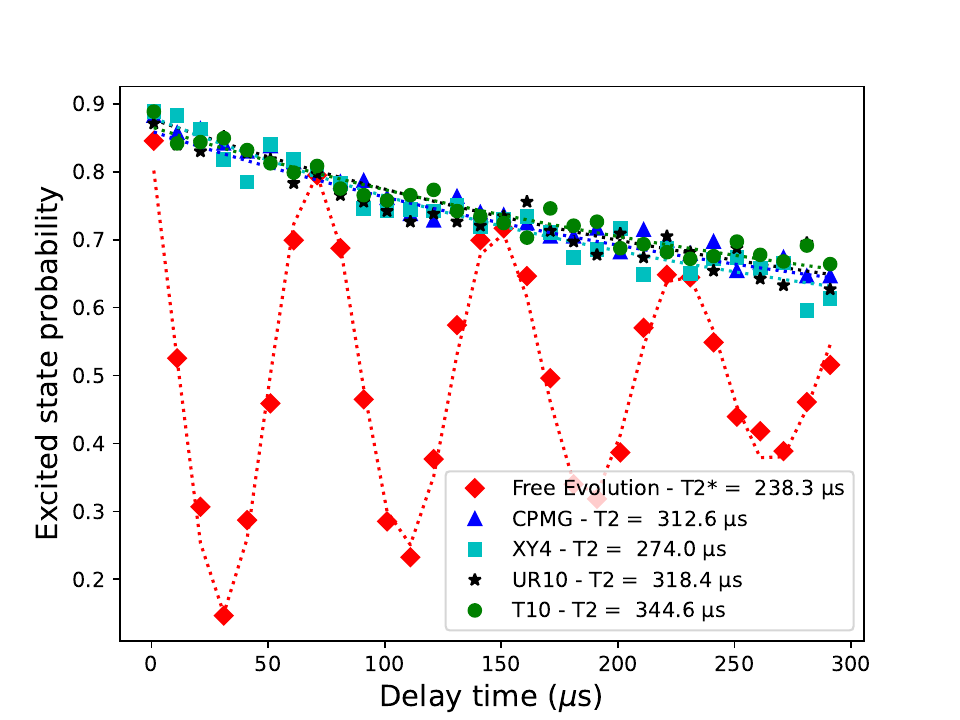}

    \includegraphics[width=\linewidth]{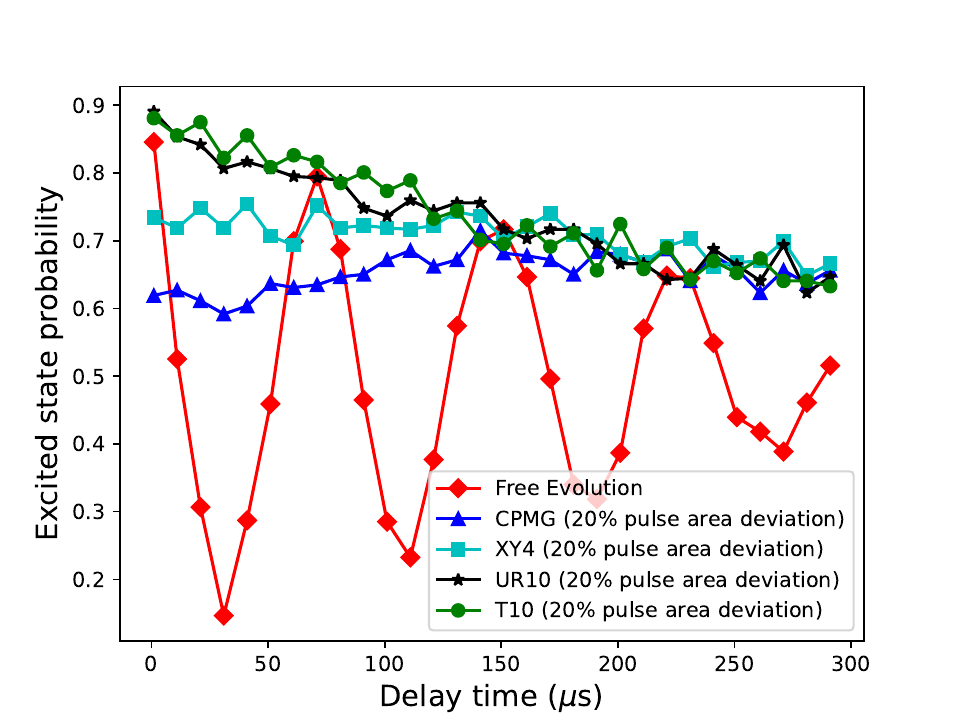}
\end{minipage}
\hfill
\begin{minipage}[h!t!]{0.48\textwidth}
    \centering
    \includegraphics[width=\linewidth]{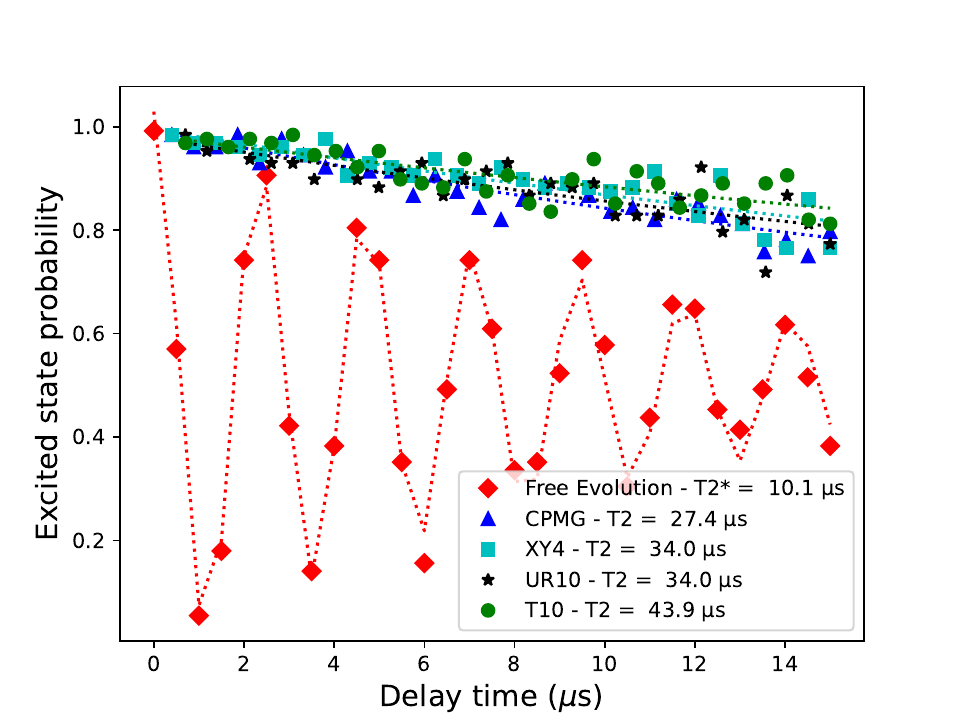}

    \includegraphics[width=\linewidth]{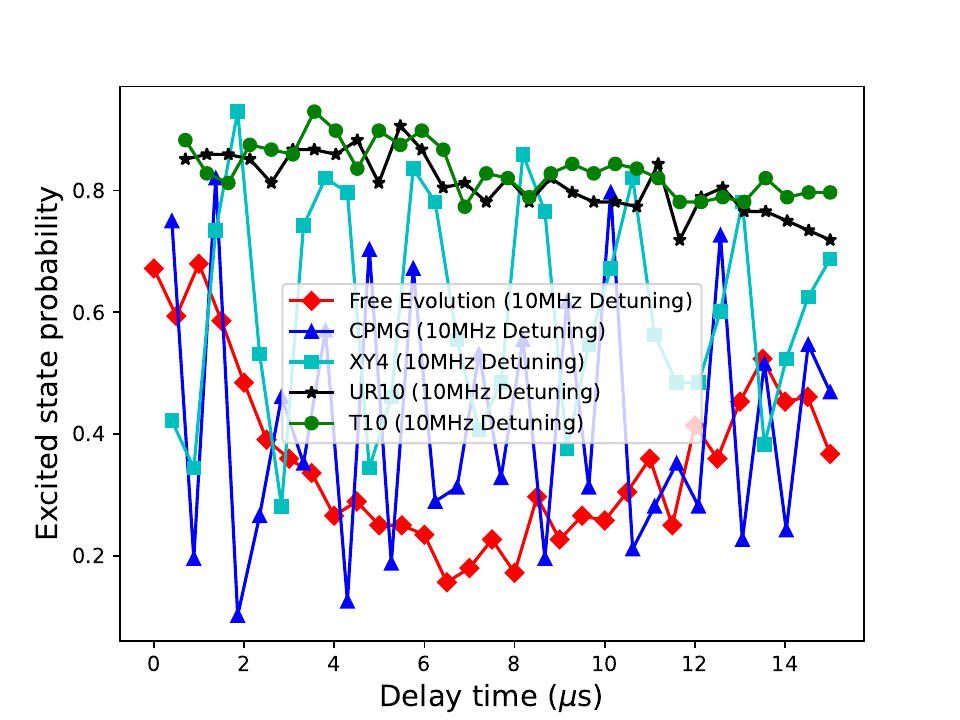}
\end{minipage}
\caption{
\oldtext{Comparison of $T2$ measurements.}
\newtext{Ramsey-coherence comparison.}
\textit{Left column:} Ramsey experiments performed on ibm{\_}torino using different dynamical decoupling sequences. Results are shown for pulses with nominal $\pi$ area (top) and for pulses with a 20\% deviation from the nominal pulse area (bottom).
\textit{Right column:} Ramsey measurements on IQM Garnet, highlighting detuning sensitivity across different dynamical decoupling sequences. Data are shown without added detuning (top) and with a 10 MHz induced detuning (bottom).
}
\label{Fig:T2_comparison}
\end{figure*}

\paragraph{Calibration data.}
The demonstration on the ibm{\_}torino device was performed
on 05/08/2025. At the time of the demonstration, the parameters of the 10th qubit of the ibm{\_}torino system were calibrated as follows: the coherence time T1 was 224 $\mu$s, the coherence time T2 was 256 $\mu$s, and the readout assignment error was 1.66\%.
The demonstration on the IQM Garnet system was conducted on 06/09/2025. At the time of the demonstration, the parameters of the 1st qubit of the IQM Garnet system were calibrated as follows: the qubit frequency was 4.4955 GHz, the T1 coherence time was 44.45 $\mu$s, the T2 (Ramsey) coherence time was 5.88 $\mu$s, the T2 (echo) coherence time was 18.09 $\mu$s, and the readout assignment error was 2.10\%.

\paragraph{Control frame and phase implementation.}
All control operations are described in the rotating frame of the qubit drive.
On the IBM Quantum processor, phase shifts are implemented as virtual $R_Z(\phi)$ gates, which correspond to instantaneous frame updates and do not introduce additional control errors or decoherence.
The exact implementation of a pulse with a phase used in our experimental demonstration is $R_Z(-\phi)R_X(\theta)R_Z(\phi)$ where \cite{RXGate,RZGate}
\begin{equation}
R_X(\theta) = \exp\left(-i\frac{\theta}{2}X \right)
= 
\begin{pmatrix}
\cos(\theta/2) & -i\sin(\theta/2) \\
-i\sin(\theta/2) & \cos(\theta/2)
\end{pmatrix}
\end{equation}
\begin{equation}
R_Z(\phi)=\exp\left(-i\frac{\phi}{2}Z \right)=\begin{pmatrix}
e^{-i\frac{\phi}{2}}&  0\\
0 & e^{i\frac{\phi}{2}}
\end{pmatrix}
\end{equation}
Therefore we get
\begin{equation}\label{rx_phi}
R_Z(-\phi)R_X(\theta)R_Z(\phi)=\begin{pmatrix}
\cos(\theta/2) & -i\sin(\theta/2)e^{i\phi} \\
-i\sin(\theta/2)e^{-i\phi} & \cos(\theta/2)
\end{pmatrix}.
\end{equation}
On the IQM Garnet processor, the phases are implemented directly at the pulse level by shifting the microwave drive phase and are embedded in the phased X rotation (PRX) gate definition \cite{IQM_PRX}
\begin{equation}
    R_{\phi}(\theta) = e^{-i (X \cos \phi - Y \sin \phi)\, \theta / 2},
\end{equation}
where $X$ and $Y$ gates in the definition are the same as the $x$ and $y$ Pauli matrices. This representation is equivalent to Eq. (\ref{rx_phi}).
In both platforms, the effective unitary implemented by a pulse with phase $\phi$ is therefore equivalent to Eq.~(\ref{Uphi}), ensuring a faithful realization of the theoretical model.

\paragraph{Implementation of pulse-area deviations.}
Systematic pulse-area errors are introduced by deliberately varying the rotation angle $\theta$ of the $R_X(\theta)$ gates away from the ideal $\pi$ value.
On ibm\_torino, this is achieved using native fractional $R_X(\theta)$ gates \cite{Frac_gate}. On IQM Garnet, the pulse area deviations are implemented by scaling the pulse amplitude while keeping the pulse duration fixed.
In both cases, this procedure produces a controlled, uniform, systematic error that is consistent across all pulses in a given sequence.

\paragraph{Remaining qubits and crosstalk.}
All experiments are performed with other qubits left idle in their ground states.
No dynamical decoupling or active cancellation is applied to spectator qubits.
The observed robustness is therefore not attributable to multi-qubit effects or crosstalk mitigation, but arises solely from the single-qubit phase structure of the \(T2n\) sequences.

\section{Ramsey Experiment}

In this demonstration, we used qubit 10 of ibm{\_}torino and qubit 1 of IQM Garnet again. The exact dates of the demonstration were 28/01/2026 and 03/02/2026, respectively. 
At the time of the demonstration, the parameters of the 10th qubit of the ibm{\_}torino system were calibrated as follows: the coherence time T1 was 232 $\mu$s, the coherence time T2 was 252 $\mu$s, and the readout assignment error was 4.6\%. The parameters of the 1st qubit of the IQM Garnet system were calibrated as follows: the qubit frequency was 4.4949 GHz, the T1 coherence time was 48 $\mu$s, the T2 (Ramsey) coherence time was 9.8 $\mu$s, the T2 (echo) coherence time was 21.1 $\mu$s, and the readout assignment error was 2.82\%.
The Ramsey experiment proceeds as follows: a $\pi/2$ pulse is applied, followed by a variable delay, a second $\pi/2$ pulse, and a final measurement. On the IBM platform, where the intrinsic detuning frequency is relatively small, an additional phase gate is inserted before the second $\pi/2$ pulse to enhance the visibility of Ramsey oscillations. The applied phase is given by $\varphi = 2\pi f t$, where $f$ is a user-defined oscillation frequency set to 100 kHz and $t$ is the delay time. On the IQM platform, an equivalent effect is obtained by directly inducing a detuning, chosen here to be 400 kHz. For the remaining experiments, a single instance of the DD sequence is applied within each delay interval.
The upper panels of Fig.~\ref{Fig:T2_comparison} show the results for free evolution and for the various DD sequences. The free-evolution data are fitted using
\begin{equation}
    P(t) = A\, e^{-t/T_{2}^{*}}\cos\!\left(2\pi B\, t\right) + 0.5
\end{equation}
while the DD data are fitted with
\begin{equation}
    P(t) = A\, e^{-t/T_{2}} + 0.5
\end{equation}
allowing extraction of the coherence times $T_{2}^{*}$ and $T_{2}$, respectively. Here, $A$ and $B$ are fit parameters.

The lower panels of Fig.~\ref{Fig:T2_comparison} present the same set of experiments as in the top panels, but with controlled imperfections introduced into the pulses forming the DD sequences. Specifically, a 20\% deviation in pulse area is applied in the left graph, while an induced detuning of 10 MHz is introduced in the right graph. 
\oldtext{In the absence of additional pulse imperfections, the T10 sequence yields the longest fitted coherence time among the sequences tested here.}
\newtext{In the absence of additional pulse imperfections, T10 yields the longest fitted coherence time among the sequences tested here. On ibm{\_}torino, the fitted values are $T_2^*=238.3~\mu$s for free evolution and $T_2=312.6$, $274.0$, $318.4$, and $344.6~\mu$s for CPMG, XY4, UR10, and T10, respectively. On IQM Garnet, the corresponding values are $T_2^*=10.1~\mu$s for free evolution and $T_2=27.4$, $34.0$, $34.0$, and $43.9~\mu$s for CPMG, XY4, UR10, and T10, respectively. Thus, T10 extends the fitted coherence time relative to free evolution by factors of approximately 1.45 and 4.35 on the two processors.}
When pulse-area errors or detuning are present, both UR10 and T10 exhibit qualitatively enhanced robustness compared to the remaining sequences.

\bibliographystyle{apsrev4-2}
\bibliography{references}